\documentclass[prd,reprint,showpacs,showkeys]{revtex4-1}
\usepackage{amssymb}
\usepackage{amsmath}
\usepackage{graphicx}
\usepackage[font={footnotesize,it}]{caption}

\pdfoutput=1

\begin{document}

\title{Quantum probes of timelike naked singularities in the weak field
regime of $f(R)$ global monopole spacetime}
\author{O. Gurtug}
\email{ozay.gurtug@emu.edu.tr}
\author{M. Halilsoy}
\email{mustafa.halilsoy@emu.edu.tr}
\author{S. Habib Mazharimousavi}
\email{habib.mazhari@emu.edu.tr}
\date{\today }

\begin{abstract}
The formation of a naked singularity in $f(R)$ global monopole spacetime is
considered in view of quantum mechanics. Quantum test fields obeying the
Klein$-$Gordon, Dirac and Maxwell equations are used to probe the classical
timelike naked singularity developed at $r=0$. We prove that the spatial
derivative operator of the fields fails to be essentially self-adjoint. As a
result, the classical timelike naked singularity formed in $f(R)$ global
monopole spacetime remains quantum mechanically singular when it is probed
with quantum fields having different spin structures. Pitelli and Letelier
(Phys. Rev. D 80, 104035, 2009) had shown that for quantum scalar ($spin$ $0$%
) probes the general relativistic global monopole singularity remains
intact. For specific modes electromagnetic ($spin$ $1$) and Dirac field ($%
spin$ $1/2$) probes, however, we show that the global monopole spacetime
behaves quantum mechanically regular. The admissibility of this singularity
is also incorporated within the Gubser's singularity conjecture.
\end{abstract}

\maketitle
\affiliation{Department of Physics, Eastern Mediterranean University, G. Magusa, north
Cyprus, Mersin 10 - Turkey}
\affiliation{Department of Physics, Eastern Mediterranean University, G. Magusa, north
Cyprus, Mersin 10 - Turkey}

\section{Introduction}

Spacetime singularities are believed to be one of the inevitable
consequences of the Einstein's theory of relativity. It describes the "end
point" or incomplete geodesics for timelike or null trajectories followed by
classical particles. \ The black hole and colliding plane wave spacetimes
are the two important branches of this theory that the nature and
characteristics of spacetime singularities are manifested. Another
intriguing one is the Big-Bang-like cosmological singularities. According to
the classical singularity classification devised by Ellis and Schmidt \cite%
{1}, curvature singularities can be grouped as \textit{scalar} and \textit{%
nonscalar. }The\textit{\ scalar }curvature\textit{\ }singularities are the
strongest ones in the sense that the spacetime becomes inextendible and all
the physical quantities, such as the gravitational field, energy density and
tidal forces, diverge at the singular point. Singularities forming at the
centre of black holes and in some colliding plane wave spacetimes are good
examples for strong \textit{scalar} curvature singularity. In black hole
spacetimes singularities located at the centre ($r=0$) is hidden by
horizon(s). In the cases where this singularity is not hidden, it is called
the \textit{naked} singularity. Whereas, the singularity occurring in the
interaction region of Bell-Szekeres solution \cite{BS} which describes the
nonlinear interaction of electromagnetic plane waves can be given as an
example to \textit{nonscalar} curvature singularity.

Naked singularity which is visible from outside needs further care as far as
the weak cosmic censorship hypothesis is concerned. It is beleived that,
naked singularity forms a threat to this hypothesis. Hence, understanding
and the resolution of naked singularities seems to be extremely important
for the deterministic nature of general relativity.

However, the scale where the singularities are forming is very small
(smaller than the Planck scale), so that the classical general relativity
methods in the resolution of the singularities are expected to be replaced
by the quantum theory of gravity. Unfortunately, there is no consistent
quantum theory of gravity yet. Since this theory is still "under
construction", the alternative methods in healing the singularities are
always attracted the attentions. String theory \cite{2,3}and loop quantum
gravity \cite{4} constitutes two major study fields in resolving
singularities. It is shown in string theory that some timelike singularities
are resolved: the orbifold, the flop, and the conifold. The flop and the
conifold occurs in the Calabi-Yau manifolds in which their resolution
involves the use of light matters such as "twisted sectors" and "wrapped
D-branes" \cite{5}(and references therein).

A rather different approach is considered in \cite{amit}\ for resolving the
timelike singularities in Reissner-Nordstr\"{o}m and negative mass
Schwarzschild solutions. In this approach, the spacetime is viewed as being
made of two parts which are naturally connected across the singularity. In
this study, it is shown that the Reissner-Nordstr\"{o}m singularity allows
for communication through the singularity and can be termed as "beam
splitter" since the transmission probability of a suitably prepared high
energy wave packet is 25\%.

Another alternative method; following the work of Wald \cite{6}, is proposed
by Horowitz and Marolf (HM)\cite{7}, which incorporates "self-adjointness"
of the spatial part of the wave operator. Hence, the classical notion of 
\textit{geodesics incompleteness} with respect to point-particle probe will
be replaced by the notion of \textit{quantum singularity} with respect to
wave probes.

The method of HM has been used successfully for other spacetimes to check
whether the classically singular spacetimes are quantum mechanically regular
or not. As an example; negative mass Schwarzschild spacetime, charged
dilatonic black hole spacetime and fundamental string spacetimes are
considered in \cite{7}. An alternative function space, namely the Sobolev
space instead of the Hilbert space, has been introduced in \cite{8}, for
analyzing the singularities within the framework of quantum mechanics. As a
result, the occurrence of timelike naked singularity in the negative mass
Schwarzschild solution is shown to be quantum mechanically regular.
Helliwell and Konkowski have studied quasiregular \cite{9},
Gal'tsov-Letelier-Tod spacetime \cite{10}, Levi-Civita spacetimes \cite%
{11,12}, and recently, they have also considered conformally static
spacetimes \cite{13,14}. Pitelli and Letelier have studied spherical and
cylindrical topological defects \cite{15}, Banados$-$Teitelboim$-$Zanelli
(BTZ) spacetimes \cite{16}, the global monopole spacetime \cite{17} and
cosmological spacetimes \cite{18}. Quantum singularities in matter coupled $%
2+1$ dimensional black hole spacetimes are considered in \cite{19}. Quantum
singularities are also considered in Lovelock theory \cite{20} and linear
dilaton black hole spacetimes \cite{21}. The occurrence of naked
singularities in a $2+1$ dimensional magnetically charged solution in
Einstein-Power-Maxwell theory have also been considered \cite{22}. Recently,
the formation of naked singularity in a model of $f(R)$ gravity is
considered in \cite{23}.

The main motivation in these studies is to understand whether these
classically singular spacetimes turn out to be quantum mechanically regular
if they are probed with quantum fields rather than classical particles.

Recently, a solution \ describing $f(R)$ global monopole in the weak field
regime has been presented in \cite{24}. This study showed that, the main
contribution of the modified theory compared to the ordinary global monopole
solution due to the Barriola and Vilenkin (BV) \cite{25} is that, in
addition to admitting double and single horizons, it admits solution without
horizon as well. And, the most important influence is seen on the nature of
the singularity that occur at $r=0$. In the case of BV, this singularity is
spacelike, whereas in the case of $f(R)$ theory, it has timelike nature.

Generally, solutions admitting black holes attracted more attention than the
solutions admitting naked singularity. Recently, the influence of the
modified theory on the thermodynamic quantities of an $f(R)$ global monopole
spacetime \cite{24} has been investigated and compared with BV spacetime in 
\cite{26}. The outcome of this investigation is that, $f(R)$ theory modifies
the thermodynamic quantities, but the shapes of curves for thermodynamic
quantities with respect to the horizon are similar to the results within the
frame of general relativity.

In this paper, we wish to investigate the occurrence of timelike naked
singularities in $f(R)$ global monopole spacetime within the context of
quantum mechanics. The singularity at $r=0$ will be probed with three
different types of quantum fields that obey Klein-Gordon, Maxwell and Dirac
equations. The singularity for the BV spacetime will also be investigated
with the spinor fields obeying Maxwell and Dirac equations.This will be the
spinor field generalization of the study performed by Pitelli and Letelier 
\cite{17} for BV spacetime.

The appearance of naked singularities are also encountered in gauged
supergravity theories. Gubser \cite{GUB} proposed a singularity conjecture
to resolve singularities in these theories in the following way.

\textbf{Conjecture}\textit{: Large curvatures in scalar coupled gravity with
four dimensional Poincare invariant solution are allowed only if the scalar
potential is bounded above in the solution.}

In this paper, the approach of Gubser will be incorporated to our analysis
briefly to display its applicability in spacetimes which do not obey
Poincare invariance.

The paper is organized as follows: In section II, we give the solution and
the spacetime structure obtained in \cite{24}. The definition of quantum
singularity is briefly reviewed in section III. Section IV is devoted for
the quantum singularity analysis of the $f(R)$ global monopole spacetime.
Three different types of waves with different spins are used to probe the
singularity. The spinor field generalization of the paper by Pitelli and
Letelier \cite{17} is given in section V. In section VI, Gubser's
singularity conjecture is used to identify if the studied curvature
singularity is \textit{bad} or \textit{good}. Finally, we give the
concluding remarks of this study in section VII.

\section{The Metric for a Global Monopole in $f(R)$ Theories and Spacetime
Structure}

\subsection{The Metric for a Global Monopole in $f(R)$ Theories.}

Recently, the metric describing the global monopole in $f\left( R\right) $
theories for the static spherically symmetric systems has been presented in
the weak field regime \cite{24}. The adopted action for such a gravitational
field coupled to matter fields in $f\left( R\right) $ theory is given by%
\begin{equation}
S=\frac{1}{2\kappa }\int d^{4}x\sqrt{-g}f(R)+S_{m},
\end{equation}%
in which $f(R)$ is an analytic function of the Ricci scalar $R$, $\kappa
=8\pi G,$ here $G$ is the Newton constant and $S_{m}$ represents the action
of the coupled matter fields given by%
\begin{equation}
S_{m}=\int d^{4}x\sqrt{-g}\mathcal{L}.
\end{equation}%
In the considered global monopole model, $\mathcal{L}$ represents the
Lagrangian density that gives the simplest global monopole model given by%
\begin{equation}
\mathcal{L}=\frac{1}{2}\partial _{\mu }\phi ^{a}\partial ^{\mu }\phi ^{a}-%
\frac{1}{4}\lambda \left( \phi ^{a}\phi ^{a}-\eta ^{2}\right) ,
\end{equation}%
in which $\lambda $ and $\eta $ are constant parameters. The global
monopole, that forms as a result of spontaneous symmetry breaking from
global $O(3)$ to $U(1),$ during the phase transitions in the early universe
is described by the self - coupling triplet of scalar fields $\phi ^{a}$ $%
\left( a=1,2,3\right) $ given by the following ansatz,%
\begin{equation}
\phi ^{a}=\eta \frac{x^{a}}{r},
\end{equation}%
with $x^{a}x^{a}=r^{2}$ and $\eta $ is a constant parameter. The adopted
metric for such a model is given by%
\begin{equation}
ds^{2}=Bdt^{2}-Adr^{2}-r^{2}\left( d\theta ^{2}+\sin ^{2}\theta d\varphi
^{2}\right) ,
\end{equation}%
where $B=B\left( r\right) $ and $A=A\left( r\right) $ are only function of $r
$. The field equation reads 
\begin{multline}
F(R)R_{\mu }^{\nu }+ \\
\left( \square F(R)-\frac{1}{2}f(R)\right) \delta _{\mu }^{\nu }-\nabla
^{\nu }\nabla _{\mu }F(R)=\kappa T_{\mu }^{\nu }
\end{multline}%
in which 
\begin{equation}
F(R)=\frac{df\left( R\right) }{dR},
\end{equation}%
\begin{equation}
\square F(R)=\frac{1}{\sqrt{-g}}\partial _{\mu }\left( \sqrt{-g}\partial
^{\mu }\right) F(R)
\end{equation}%
and%
\begin{equation}
\nabla ^{\nu }\nabla _{\mu }F(R)=g^{\alpha \nu }\left[ \left( F(R)\right)
_{,\mu ,\alpha }-\Gamma _{\mu \alpha }^{m}\left( F(R)\right) _{,m}\right] .
\end{equation}%
In Eq. (6) $T_{\mu }^{\nu }$ represents minimally coupled energy $-$
momentum tensor of the matter field whose non-zero components are given by%
\begin{multline}
T_{0}^{0}=T_{r}^{r}= \\
-\frac{8\pi G\eta ^{2}+3GM\psi _{0}}{r^{2}}+\frac{3-16\pi G\eta ^{2}}{r}\psi
_{0}+3\psi _{0}^{2}.
\end{multline}%
Furthermore, the trace of the field equation (6) reads%
\begin{equation}
F(R)R+3\square F(R)-2f(R)=\kappa T,
\end{equation}%
with $T=T_{\mu }^{\mu }.$ With reference to the paper \cite{24}, the
solution to the field equations was obtained in the weak field regime which
assumes the metric function in the form of $B=1+b(r)$ and \ $A=1+a(r)$ with
the property that $\left\vert a\left( r\right) \right\vert $ and $\left\vert
b\left( r\right) \right\vert $ smaller than unity. As a consequence of a
weak field regime, the considered model of $f(R)$ theory corresponds to a
small correction on standard general relativity in such a way that, $%
F(R(r))=1+\psi (r)$ with $\psi (r)\ll 1.$ Explicit form of $f(R)$ is given
in \cite{24} (Eq. 42 in \cite{24}). Hence, $F(R(r))=1$ corresponds to the
standard general relativity. Employing these conditions in the field
equations yields $\psi (r)=\psi _{0}r$ and resulting metric function with
global monopole is found to be%
\begin{equation}
B=A^{-1}=1-8\pi G\eta ^{2}-\frac{2GM}{r}-\psi _{0}r,
\end{equation}%
where $M$ is the mass parameter and $\psi _{0}$ is a very small parameter (
since $\psi _{0}r<<1$) that measures the deviation from the standard general
relativity. As stated in \cite{24}, for a typical Grand Unified Theory the
parameter $\eta $ is in the order of $10^{16}$ GeV. Hence, $8\pi G\eta
^{2}\approx 10^{-5}.$ Note that one can recover the result of BV if $\psi
_{0}=0.$ It is known that, the global monopole solution obtained by BV has
one horizon only and the nature of the singularity at $r=0$ is spacelike.

\subsection{The Spacetime Structure}

The structure of the solution obtained in \cite{24} and given in Eq. (12),
has remarkable features that deserves to be investigated in detail. \ The
obtained solution admit black holes with inner and outer horizons. To find
the location of the horizon, we prefer to write the metric component $g_{tt}$
in the following form%
\begin{equation}
B=-\frac{\psi _{0}}{r}\left( r-r_{+}\right) \left( r-r_{-}\right)
\end{equation}%
where $r_{+}$ and $r_{-}$ denote the outer and inner horizons respectively
and given by%
\begin{equation}
r_{\pm }=\frac{\alpha \pm \sqrt{\alpha ^{2}-8\psi _{0}GM}}{2\psi _{0}},\text{
\ \ }\alpha =1-8\pi G\eta ^{2}.
\end{equation}%
The Kretschmann scalar which indicates the formation of curvature
singularity for the $f(R)$ global monopole is given by%
\begin{multline}
\mathcal{K}=\frac{4}{r^{6}}\left\{ 2\psi _{0}^{2}r^{4}+\left( 16\psi _{0}\pi
G\eta ^{2}\right) r^{3}\right. + \\
\left. \left( 8\pi G\eta ^{2}\right) ^{2}r^{2}+\left( 32\pi G^{2}M\eta
^{2}\right) r+12GM^{2}\right\} .
\end{multline}%
It is evident that $r=0$ is a typical central curvature singularity that is
peculiar to the spherically symmetric systems. In order to find the nature
or the character of the singularity at $r=0$ for the $f(R)$ global monopole,
we perform conformal compactification. The conformal radial or tortoise
coordinate is given by 
\begin{multline}
r_{\ast }=\int \frac{dr}{B}= \\
-\frac{1}{\psi _{0}\left( r_{+}-r_{-}\right) }\left\{ r_{+}\ln \left\vert
r-r_{+}\right\vert -r_{-}\ln \left\vert r-r_{-}\right\vert \right\} .
\end{multline}%
The retarded and advanced coordinates are defined as $u=t-r_{\ast }$ and $%
v=t+r_{\ast }$ respectively. Defining the Kruskal coordinates as 
\begin{eqnarray}
u^{^{\prime }} &=&\exp \left( \frac{\psi _{0}\left( r_{+}-r_{-}\right) }{%
2r_{-}}u\right) ,\text{ \ } \\
v^{^{\prime }} &=&\exp \left( -\frac{\psi _{0}\left( r_{+}-r_{-}\right) }{%
2r_{-}}v\right) \text{,}
\end{eqnarray}%
the metric can be written as%
\begin{equation}
ds^{2}=\frac{4r_{-}^{2}\left( r-r_{+}\right) ^{\frac{r_{+}+r_{-}}{r_{-}}}}{%
\psi _{0}r\left( r_{+}-r_{-}\right) ^{2}}du^{^{\prime }}dv^{^{\prime
}}-r^{2}\left( d\theta ^{2}+\sin ^{2}\theta d\varphi ^{2}\right) ,
\end{equation}%
and 
\begin{equation}
u^{^{\prime }}v^{^{\prime }}=\left( r-r_{-}\right) \left( r-r_{+}\right)
^{-r_{+}/r_{-}}.
\end{equation}%
In order to bring infinity into a finite coordinate, we define 
\begin{eqnarray}
u^{^{\prime \prime }} &=&\arctan u^{^{\prime }},\text{\ \ }0<u^{^{\prime
\prime }}<\pi /2,\text{\ } \\
\text{\ \ \ \ \ \ }v^{^{\prime \prime }} &=&\arctan v^{^{\prime }},\text{ \
\ }0<v^{^{\prime \prime }}<\pi /2.\text{\ }
\end{eqnarray}%
The corresponding Carter - Penrose diagrams for the following three possible
cases are plotted and given in figures. The singularity located at $r=0$ is
shown vertically on the Carter-Penrose diagram which indicates timelike
character.

There are three possible cases to be investigated.

\subsubsection{Case 1: When $\protect\alpha ^{2}>8\protect\psi _{0}GM.$}

The metric function, $B(r)=\alpha -\frac{2GM}{r}-\psi _{0}r,$ admits two
positive roots $r_{+}$ and $r_{-},$ indicating the location of the outer and
inner horizons of a black hole. The Penrose diagram for this case is shown
in Fig.1.

\begin{figure}[tbp]
\includegraphics[width=60mm,scale=0.7]{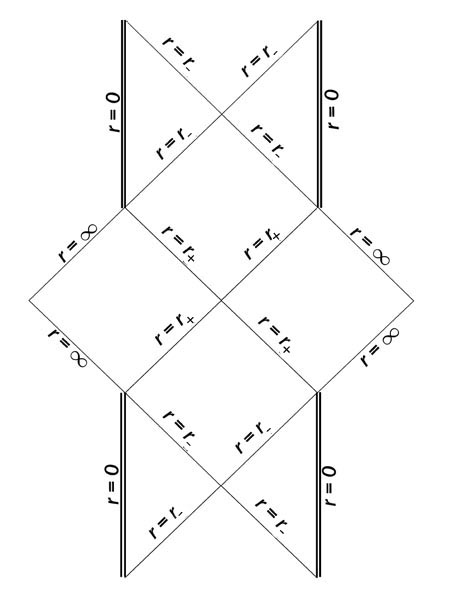}
\caption{Carter - Penrose diagram of the $f(R)$ global monopole spacetime
with inner $r_{-}$ and outer $r_{+}$ horizons. Timelike singularity is at $%
r=0$. }
\end{figure}

\subsubsection{Case 2: When $\protect\alpha ^{2}=8\protect\psi _{0}GM.$}

The metric function, $B(r)=\alpha -\frac{2GM}{r}-\psi _{0}r,$ admits one
horizon only. It can be interpreted as the extreme black hole. The Penrose
diagram of this case is given in Fig. 2. Recently, the thermodynamic
properties of the black hole solutions of $f(R)$ global monopole is
investigated and presented in \cite{25}.

\begin{figure}[tbp]
\includegraphics[width=40mm,scale=0.7]{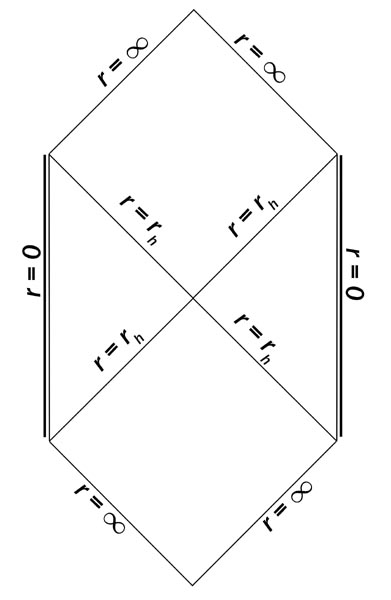}
\caption{Carter - Penrose diagram of $f(R)$ global monopole spacetime with a
single horizon at $r=r_{h}.$}
\end{figure}

\subsubsection{Case 3: When $\protect\alpha ^{2}<8\protect\psi _{0}GM.$}

In this case, the metric function, $B(r)=\alpha -\frac{2GM}{r}-\psi _{0}r$,
does not admit real roots. Hence, the solution in this particular case is
not a black hole solution and the singularity at $r=0$ becomes timelike
naked singularity, as depicted in the Penrose diagram in Fig. 3. The choice
of the parameters of the $f(R)$ global monopole metric results with timelike
naked singularity at $r=0$ or black hole solutions with one or two horizons.
These results seem to show that the small correction to the standard general
relativity produces significant changes on the spacetime structure of the BV
metric obtained by Barriola and Vilenkin.

In this paper, we are aiming to investigate this singularity within the
context of quantum mechanics. This classically singular spacetime will be
probed with quantum waves obeying the Klein-Gordon, Maxwell and Dirac
equations to check whether the timelike naked singularity is smoothed out or
not. 
\begin{figure}[tbp]
\includegraphics[width=30mm,scale=0.7]{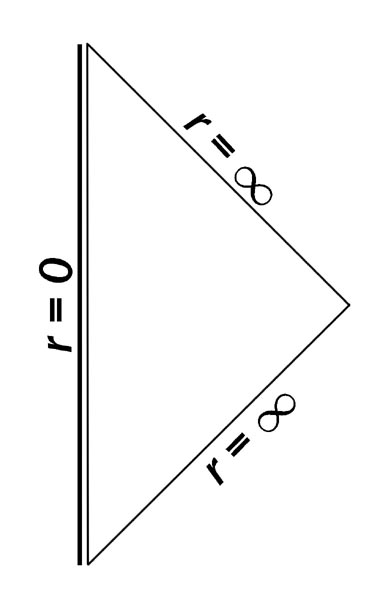}
\caption{Carter - Penrose diagram for $f(R)$ global monopole spacetime
without a horizon in which $r=0$ is a naked timelike singularity. }
\end{figure}

\subsection{The Description of the $f(R)$ Global Monopole Spacetime in a
Newman-Penrose (NP) Formalism}

The $f(R)$ global monopole metric is investigated with the Newman-Penrose
(NP) formalism, in order to clarify the contribution of the $f(R)$ gravity.
\ The set of proper null tetrads $1-forms$ is given by%
\begin{eqnarray}
l &=&dt-\frac{dr}{B(r)}, \\
n &=&\frac{1}{2}\left( B(r)dt+dr\right) , \\
m &=&-\frac{r}{\sqrt{2}}\left( d\theta +i\sin \theta d\varphi \right) . \\
\bar{m} &=&-\frac{r}{\sqrt{2}}\left( d\theta -i\sin \theta d\varphi \right)
\end{eqnarray}%
The non-zero spin coefficients in these tetrads are%
\begin{eqnarray}
\beta &=&-\alpha =\frac{\cot \theta }{2\sqrt{2}r},\text{ \ \ }\rho =-\frac{1%
}{r},\text{ \ } \\
\text{\ }\mu &=&-\frac{B}{2r},\text{ \ \ }\gamma =\frac{1}{4}\frac{dB}{dr}.
\end{eqnarray}%
As a result, we obtain the Weyl and the Ricci scalars as%
\begin{equation}
\Psi _{2}=-\frac{3GM+4\pi G\eta ^{2}r}{3r^{3}},
\end{equation}%
\begin{equation}
\phi _{11}=\frac{8\pi G\eta ^{2}+\psi _{0}r}{4r^{2}},
\end{equation}%
\begin{equation}
\Lambda =\frac{8\pi G\eta ^{2}+3\psi _{0}r}{12r^{2}},
\end{equation}%
so that the spacetime is Petrov type$-D$. The parameter $\psi _{0}$
representing the contribution of $f(R)$gravity is seen to effect only the
Ricci components, leaving the mass term $\Psi _{2}$ of an ordinary global
monopole unchanged.

\section{Quantum Singularities}

Horowitz and Marolf (HM) \cite{7}, by developing the pioneering work of Wald 
\cite{6}, have proposed a prescription which involves the use of quantum
particles/waves to judge whether the classical timelike curvature
singularities occurring in static spacetimes are smoothed out quantum
mechanically or not. According to HM, the singular character of the
spacetime is defined as the ambiguity in the evolution of the wave
functions. That is to say, the singular character is determined in terms of
the ambiguity when attempting to find a self-adjoint extension of the
spatial part of the wave operator to the entire Hilbert space. If the
extension is unique, it is said that the space is quantum mechanically
regular. A brief review now follows:

Consider a static spacetime $\left( M,g_{\mu \nu }\right) $\ with a timelike
Killing vector field $\xi ^{\mu }$. Let $t$ denote the Killing parameter and 
$\Sigma $\ denote a static slice. The Klein-Gordon equation in this space is

\begin{equation}
\left( \nabla ^{\mu }\nabla _{\mu }-m^{2}\right) \psi =0.
\end{equation}%
This equation can be written in the form

\begin{equation}
\frac{\partial ^{2}\psi }{\partial t^{2}}=\sqrt{f}D^{i}\left( \sqrt{f}%
D_{i}\psi \right) -fm^{2}\psi =-A\psi ,
\end{equation}%
in which $f=-\xi ^{\mu }\xi _{\mu }$ and $D_{i}$ is the spatial covariant
derivative on $\Sigma $. The Hilbert space $\mathcal{H}$, $\left(
L^{2}\left( \Sigma \right) \right) $\ is the space of square integrable
functions on $\Sigma $. The domain of an operator $A,$ $D(A),$ is taken in
such a way that it does not enclose the spacetime singularities. An
appropriate set is $C_{0}^{\infty }\left( \Sigma \right) $, the set of
smooth functions with compact support on $\Sigma $. The operator $A$ is
real, positive and symmetric; therefore, its self-adjoint extensions always
exist. If \ it has a unique extension $A_{E},$ then $A$ is called
essentially self-adjoint \cite{27,28,29}. Accordingly, the Klein-Gordon
equation for a free particle satisfies

\begin{equation}
i\frac{d\psi}{dt}=\sqrt{A_{E}}\psi,
\end{equation}
with the solution

\begin{equation}
\psi \left( t\right) =\exp \left[ -it\sqrt{A_{E}}\right] \psi \left(
0\right) .
\end{equation}%
If $A$ is not essentially self-adjoint, the future time evolution of the
wave function (35) is ambiguous. Then the HM criterion defines the spacetime
as quantum mechanically singular. However, if there is only a single
self-adjoint extension, the operator $A$ is said to be\ essentially
self-adjoint and the quantum evolution described by Eq. (35) is uniquely
determined by the initial conditions. According to the HM criterion, this
spacetime is said to be quantum mechanically non-singular. In order to
determine the number of self-adjoint extensions, the concept of deficiency
indices is used. The deficiency subspaces $N_{\pm }$ are defined by (see
Ref. \cite{8} for a detailed mathematical background)

\begin{gather}
N_{+}=\{\psi \in D(A^{\ast }),\text{\ }A^{\ast }\psi =Z_{+}\psi ,\text{ }%
ImZ_{+}>0\} \\
\text{ with dimension }n_{+}  \notag
\end{gather}%
\begin{gather}
N_{-}=\{\psi \in D(A^{\ast }),\text{ }A^{\ast }\psi =Z_{-}\psi ,\text{ }%
ImZ_{-}<0\} \\
\text{ with dimension }n_{-}  \notag
\end{gather}%
The dimensions $\left( \text{ }n_{+},n_{-}\right) $ are the deficiency
indices of the operator $A$. The indices $n_{+}(n_{-})$ are completely
independent of the choice of $Z_{+}(Z_{-})$ depending only on whether or not 
$Z$ lies in the upper (lower) half complex plane. Generally one takes $%
Z_{+}=i\lambda $ and $Z_{-}=-i\lambda $ , where $\lambda $ is an arbitrary
positive constant necessary for dimensional reasons. The determination of
deficiency indices is then reduced to counting the number of solutions of $%
A^{\ast }\psi =Z\psi $; (for $\lambda =1$),

\begin{equation}
A^{\ast }\psi \pm i\psi =0
\end{equation}%
that belong to the Hilbert space $\mathcal{H}$. If there are no square
integrable solutions ( i.e. $n_{+}=n_{-}=0)$, the operator $A$ possesses a
unique self-adjoint extension and is essentially self-adjoint. Consequently,
the way to find a sufficient condition for the operator $A$ to be
essentially self-adjoint is to investigate the solutions satisfying Eq. (38)
that do not belong to the Hilbert space.

\section{QUANTUM SINGULARITIES IN $f(R)$ GLOBAL MONOPOLE SPACETIME}

\subsection{Klein-Gordon Fields}

The massive Klein-Gordon equation for a scalar particle with mass $m$ can be
written as

\begin{equation}
\left( g^{-1/2}\partial _{\mu }\left[ g^{1/2}g^{\mu \nu }\partial _{\nu }%
\right] -m^{2}\right) \psi =0.
\end{equation}%
For the metric (5), the Klein-Gordon equation can be splitted into a time
and spatial part and written as

\begin{multline}
\frac{\partial ^{2}\psi }{\partial t^{2}}=-B\left\{ B\frac{\partial ^{2}\psi 
}{\partial r^{2}}+\frac{1}{r^{2}}\frac{\partial ^{2}\psi }{\partial \theta
^{2}}+\frac{1}{r^{2}\sin ^{2}\theta }\frac{\partial ^{2}\psi }{\partial
\varphi ^{2}}+\right. \\
\left. \frac{\cot \theta }{r^{2}}\frac{\partial \psi }{\partial \theta }%
+\left( \frac{2B}{r}+B^{^{\prime }}\right) \frac{\partial \psi }{\partial r}%
\right\} +Bm^{2}\psi .
\end{multline}%
In analogy with Eq. (33), the spatial operator $A$ for the massless case is

\begin{multline}
\emph{A}=B\left\{ B\frac{\partial ^{2}}{\partial r^{2}}+\frac{1}{r^{2}}\frac{%
\partial ^{2}}{\partial \theta ^{2}}+\frac{1}{r^{2}\sin ^{2}\theta }\frac{%
\partial ^{2}}{\partial \varphi ^{2}}\right. + \\
\left. \frac{\cot \theta }{r^{2}}\frac{\partial }{\partial \theta }+\left( 
\frac{2B}{r}+B^{^{\prime }}\right) \frac{\partial }{\partial r}\right\} ,
\end{multline}%
and the equation to be solved is $\left( \emph{A}^{\ast }\pm i\right) \psi
=0.$Using separation of variables, $\psi =R\left( r\right) Y_{l}^{m}\left(
\theta ,\varphi \right) $, we get the radial part of Eq. (38) as

\begin{equation}
R^{\prime \prime }+\frac{\left( r^{2}B\right) ^{^{\prime }}}{r^{2}B}%
R^{\prime }+\left( \frac{-l\left( l+1\right) }{r^{2}B}\pm \frac{i}{B^{2}}%
\right) R=0,
\end{equation}%
whose solutions represents spin $0$ bosonic waves and a prime denotes the
derivative with respect to $r$. The spatial operator $A$ is essentially self
adjoint if neither of two solutions of Eq. (42) is square integrable over
all space $L^{2}(0,\infty )$. Because of the complexity in finding exact
analytic solution to Eq. (42), we study the behavior of $R\left( r\right) $
near $r\rightarrow \infty $ and $r\rightarrow 0.$

\subsubsection{The case of r$\rightarrow \infty $}

The case $r\rightarrow \infty $ is topologically different compared to the
analysis for ordinary global monopole solutions reported in \cite{17}. The
asymptotic behavior of the $f(R)$ global monopole metric when $r\rightarrow
\infty $ is not conical and given by%
\begin{equation}
ds^{2}\simeq -(\alpha -\psi _{0}r)dt^{2}+\frac{dr^{2}}{\left( \alpha -\psi
_{0}r\right) }+r^{2}\left( d\theta ^{2}+\sin ^{2}\theta d\varphi ^{2}\right)
.
\end{equation}%
For the above metric, the radial equation (42),\ for $r\rightarrow \infty $
becomes,%
\begin{equation}
R^{\prime \prime }\pm \frac{i}{\left( \alpha -\psi _{0}r\right) }R=0,
\end{equation}%
whose solution is%
\begin{multline}
R_{\pm }=C_{1}\sqrt{\alpha -\psi _{0}r}J_{1}\left[ \left( \pm 1+i\right) 
\sqrt{2}\sqrt{\frac{\alpha -\psi _{0}r}{\psi _{0}^{2}}}\right] + \\
C_{2}\sqrt{\alpha -\psi _{0}r}N_{1}\left[ \left( \pm 1+i\right) \sqrt{2}%
\sqrt{\frac{\alpha -\psi _{0}r}{\psi _{0}^{2}}}\right] ,
\end{multline}%
where $C_{1}$ and $C_{2}$ are arbitrary integration constants, $J_{1}$ and $%
N_{1}$ are the first and second kind Bessel functions. The square
integrability of the above solution for each sign $\pm $ is checked by
calculating the squared norm of the above solution in which the function
space on each $t=$ constant hypersurface $\Sigma $ is defined as $\mathcal{H=%
}\{R\mid \left\Vert R\right\Vert <\infty \}.$ The squared norm for the
metric (43) is given by,

\begin{equation}
\left\Vert R\right\Vert ^{2}=\int_{r}^{\infty }\frac{\left\vert R_{\pm
}\left( r\right) \right\vert ^{2}r^{2}}{(\alpha -\psi _{0}r)}dr.
\end{equation}%
Our calculation has revealed that the obtained solution at infinity fails to
satisfy square integrability condition i.e. $\left\Vert R\right\Vert
^{2}\rightarrow \infty $ . Hence, the solution at infinity does not belong
to the Hilbert space. \ 

\subsubsection{The case of r$\rightarrow 0$}

The approximate metric near the origin is Schwarzschild like and given by%
\begin{multline}
ds^{2}\simeq -(\alpha -\frac{2GM}{r})dt^{2}+ \\
\frac{dr^{2}}{(\alpha -\frac{2GM}{r})}+r^{2}\left( d\theta ^{2}+\sin
^{2}\theta d\varphi ^{2}\right) .
\end{multline}

The radial equation (42), for the above metric reduces to

\begin{equation}
R^{\prime \prime }-\frac{\beta }{r}R=0,
\end{equation}%
in which $\beta =\frac{l\left( l+1\right) }{2GM}$, and the solution is
obtained in terms of first and second kind of Bessel's functions and given by

\begin{equation}
R=C_{3}\sqrt{r}J_{1}\left( 2\sqrt{\beta r}\right) +C_{4}\sqrt{r}N_{1}\left( 2%
\sqrt{\beta r}\right)
\end{equation}%
where $C_{3}$\ and $C_{4}$ are arbitrary integration constants. The square
integrability of the above solution is checked by calculating the squared
norm for the metric (47) which is given by,

\begin{equation}
\parallel R\parallel ^{2}=\int_{0}^{\text{constant}}\frac{\left\vert
R\right\vert ^{2}r^{2}}{(\alpha -\frac{2GM}{r})}dr<\infty
\end{equation}%
which is always square integrable near $r=0.$ Consequently, the spatial
operator $A$ is not square integrable over all space $L^{2}(0,\infty )$ and
therefore, it is not essentially self-adjoint. Hence, the classical
singularity at $r=0$ remains quantum mechanically singular when probed with
fields obeying the Klein-Gordon equation.

In the next subsections, the singularity will be probed with spinorial
fields obeying Maxwell and Dirac equations. We prefer to use same method and
terminology reported in \cite{23}.

\subsection{Maxwell Fields}

The Newman-Penrose formalism will be used to find the source-free Maxwell
fields propagating in the space of $f(R)$ global monopole spacetime. The
four coupled source-free Maxwell equations for electromagnetic fields in the
Newman-Penrose formalism is given by%
\begin{eqnarray}
D\phi _{1}-\bar{\delta}\phi _{0} &=&\left( \pi -2\alpha \right) \phi
_{0}+2\rho \phi _{1}-\kappa \phi _{2}, \\
\delta \phi _{2}-\Delta \phi _{1} &=&-\nu \phi _{0}+2\mu \phi _{1}+\left(
\tau -2\beta \right) \phi _{2}, \\
\delta \phi _{1}-\Delta \phi _{0} &=&\left( \mu -2\gamma \right) \phi
_{0}+2\tau \phi _{1}-\sigma \phi _{2}, \\
D\phi _{2}-\bar{\delta}\phi _{1} &=&-\lambda \phi _{0}+2\pi \phi _{1}+\left(
\rho -2\epsilon \right) \phi _{2},
\end{eqnarray}%
where $\phi _{0},$ $\phi _{1}$ and $\phi _{2}$ are the Maxwell spinors, $%
\epsilon ,\rho ,\pi ,\alpha ,\mu ,\gamma ,\beta $ and $\tau $ are the spin
coefficients to be found and the bar denotes complex conjugation. The null
tetrad vectors for the metric (5) are defined by%
\begin{eqnarray}
l^{a} &=&\left( \frac{1}{B},1,0,0\right) , \\
n^{a} &=&\left( \frac{1}{2},-\frac{B}{2},0,0\right) , \\
m^{a} &=&\frac{1}{\sqrt{2}}\left( 0,0,\frac{1}{r},\frac{i}{r\sin \theta }%
\right) . \\
\bar{m}^{a} &=&\frac{1}{\sqrt{2}}\left( 0,0,\frac{1}{r},\frac{-i}{r\sin
\theta }\right)
\end{eqnarray}%
The directional derivatives in the Maxwell's equations are defined by $%
D=l^{a}\partial _{a},\Delta =n^{a}\partial _{a}$ and $\delta =m^{a}\partial
_{a}.$ We define operators in the following way by assuming $\phi _{\alpha
}= $ $\phi _{\alpha }(r,\theta )e^{i(\omega t+m\varphi )}$ ($\alpha =0,1,2$)

\begin{eqnarray}
\mathbf{D}_{0} &=&D, \\
\mathbf{D}_{0}^{\dagger } &=&-\frac{2}{B}\Delta , \\
\mathbf{L}_{0}^{\dagger } &=&\sqrt{2}r\text{ }\delta \text{ and }\mathbf{L}%
_{1}^{\dagger }=\mathbf{L}_{0}^{\dagger }+\frac{\cot \theta }{2}, \\
\mathbf{L}_{0} &=&\sqrt{2}r\text{ }\bar{\delta}\text{ and }\mathbf{L}_{1}=%
\mathbf{L}_{0}+\frac{\cot \theta }{2}.
\end{eqnarray}%
The non-zero spin coefficients are given in Eq.s (27-28). The Maxwell
spinors are defined by \cite{30} 
\begin{eqnarray}
\phi _{0} &=&F_{13}=F_{\mu \nu }l^{\mu }m^{\nu } \\
\phi _{1} &=&\frac{1}{2}\left( F_{12}+F_{43}\right) =\frac{1}{2}F_{\mu \nu
}\left( l^{\mu }n^{\nu }+\overline{m}^{\mu }m^{\nu }\right) , \\
\phi _{2} &=&F_{42}=F_{\mu \nu }\overline{m}^{\mu }n^{\nu },
\end{eqnarray}%
where $F_{ij}\left( i,j=1,2,3,4\right) $ and $F_{\mu \nu }\left( \mu ,\nu
=0,1,2,3\right) $ are the components of the Maxwell tensor in the tetrad and
tensor bases, respectively. Substituting Eq.s (59-62) into the Maxwell's
equations together with non-zero spin coefficients, the Maxwell equations
become

\begin{gather}
\left( \mathbf{D}_{0}+\frac{2}{r}\right) \phi _{1}-\frac{1}{r\sqrt{2}}%
\mathbf{L}_{1}\phi _{0}=0, \\
\left( \mathbf{D}_{0}+\frac{1}{r}\right) \phi _{2}-\frac{1}{r\sqrt{2}}%
\mathbf{L}_{0}\phi _{1}=0, \\
\frac{B}{2}\left( \mathbf{D}_{0}^{\dagger }+\frac{B^{^{\prime }}}{B}+\frac{1%
}{r}\right) \phi _{0}+\frac{1}{r\sqrt{2}}\mathbf{L}_{0}^{\dagger }\phi
_{1}=0, \\
\frac{B}{2}\left( \mathbf{D}_{0}^{\dagger }+\frac{2}{r}\right) \phi _{1}+%
\frac{1}{r\sqrt{2}}\mathbf{L}_{1}^{\dagger }\phi _{2}=0.
\end{gather}%
The equations\ above will become more tractable if the variables are changed
to

\begin{equation}
\Phi _{0}=\phi _{0},\text{ \ }\Phi _{1}=\sqrt{2}r\phi _{1},\text{ \ }\Phi
_{2}=2r^{2}\phi _{2}.
\end{equation}%
Then, we have%
\begin{gather}
\left( \mathbf{D}_{0}+\frac{1}{r}\right) \Phi _{1}-\mathbf{L}_{1}\Phi _{0}=0,
\\
\left( \mathbf{D}_{0}-\frac{1}{r}\right) \Phi _{2}-\mathbf{L}_{0}\Phi _{1}=0,
\\
r^{2}B\left( \mathbf{D}_{0}^{\dagger }+\frac{B^{^{\prime }}}{B}+\frac{1}{r}%
\right) \Phi _{0}+\mathbf{L}_{0}^{\dagger }\Phi _{1}=0, \\
r^{2}B\left( \mathbf{D}_{0}^{\dagger }+\frac{1}{r}\right) \Phi _{1}+\mathbf{L%
}_{1}^{\dagger }\Phi _{2}=0.
\end{gather}%
The commutativity of the operators $\mathbf{L}$ and $\mathbf{D}$ enables us
to eliminate each $\Phi _{i}$ from above equations, and hence we have%
\begin{gather}
\left[ \mathbf{L}_{0}^{\dagger }\mathbf{L}_{1}+r^{2}B\left( \mathbf{D}_{0}+%
\frac{B^{^{\prime }}}{B}+\frac{3}{r}\right) \right. \times  \notag \\
\left. \left( \mathbf{D}_{0}^{\dagger }+\frac{B^{^{\prime }}}{B}+\frac{1}{r}%
\right) \right] \Phi _{0}\left( r,\theta \right) =0,
\end{gather}%
\begin{equation}
\left[ \mathbf{L}_{1}\mathbf{L}_{0}^{\dagger }+r^{2}B\left( \mathbf{D}%
_{0}^{\dagger }+\frac{B^{^{\prime }}}{B}+\frac{1}{r}\right) \left( \mathbf{D}%
_{0}+\frac{1}{r}\right) \right] \Phi _{1}\left( r,\theta \right) =0.
\end{equation}%
\begin{equation}
\left[ \mathbf{L}_{0}\mathbf{L}_{1}^{\dagger }+r^{2}B\left( \mathbf{D}%
_{0}^{\dagger }+\frac{1}{r}\right) \left( \mathbf{D}_{0}-\frac{1}{r}\right) %
\right] \Phi _{2}\left( r,\theta \right) =0,
\end{equation}%
The variables $r$ and $\theta $ can be separated by assuming a separable
solution in the form of%
\begin{eqnarray}
\Phi _{0}\left( r,\theta \right) &=&R_{0}\left( r\right) \Theta _{0}\left(
\theta \right) ,\text{ \ \ } \\
\Phi _{1}\left( r,\theta \right) &=&R_{1}\left( r\right) \Theta _{1}\left(
\theta \right) ,\text{ \ \ \ \ } \\
\Phi _{2}\left( r,\theta \right) &=&R_{2}\left( r\right) \Theta _{2}\left(
\theta \right) .
\end{eqnarray}%
The separation constants for Eq. (75) and Eq. (76) are the same, because $%
\mathbf{L}_{n}=-\mathbf{L}_{n}^{\dagger }\left( \pi -\theta \right) ,$ or,
in other words, the operator $\mathbf{L}_{0}^{\dagger }\mathbf{L}_{1}$
acting on $\Theta _{0}\left( \theta \right) $ is the same as the operator $%
\mathbf{L}_{0}\mathbf{L}_{1}^{\dagger }$ acting on $\Theta _{2}\left( \theta
\right) $ if we replace $\theta $ by $\pi -\theta $. However, for Eq. (77)
we will assume another separation constant. Furthermore, by defining $%
R_{0}\left( r\right) =\frac{f_{0}(r)}{rB\left( r\right) }$, $R_{1}(r)=\frac{%
f_{1}\left( r\right) }{r}$ and $R_{2}(r)=\frac{f_{2}\left( r\right) }{r}$,
the radial equations can be written as 
\begin{multline}
f_{0}^{^{\prime \prime }}(r)+\frac{2}{r}f_{0}^{^{\prime }}(r)+ \\
\left[ -i\omega \left( \frac{2}{rB}-\frac{B^{^{\prime }}}{B^{2}}\right) +%
\frac{\omega ^{2}}{B^{2}}-\frac{\epsilon ^{2}}{r^{2}B}\right] f_{0}(r)=0,
\end{multline}%
\begin{equation}
f_{1}^{^{\prime \prime }}(r)+\frac{B^{^{\prime }}}{B}f_{1}^{^{\prime }}(r)+%
\left[ \frac{\omega ^{2}}{B^{2}}-\frac{\eta ^{2}}{r^{2}B}\right] f_{1}(r)=0,
\end{equation}%
\begin{multline}
f_{2}^{^{\prime \prime }}(r)-\frac{2}{r}f_{2}^{^{\prime }}(r)+ \\
\left[ i\omega \left( \frac{2}{rB}-\frac{B^{^{\prime }}}{B^{2}}\right) +%
\frac{\omega ^{2}}{B^{2}}-\frac{\epsilon ^{2}}{r^{2}B}\right] f_{2}(r)=0,
\end{multline}%
where $\epsilon $ and $\eta $ are the separability constants and $\omega $
denotes the frequency of the photon wave.

The definition of the quantum singularity for Maxwell fields will be the
same as for the Klein$-$Gordon fields. Here, since we have three equations
governing the dynamics of the photon waves, the unique self-adjoint
extension condition on the spatial part of the Maxwell operator should be
examined for each of the three equations for all space.

\subsubsection{For the case $r\rightarrow \infty $}

The corresponding metric is given in Eq. (43). Hence, the radial parts of
the Maxwell equations, (81) , (82) and (83), become%
\begin{eqnarray}
f_{0}^{^{\prime \prime }}(r)+\frac{\omega \left( \omega -i\varphi
_{0}\right) }{\left( \alpha -\psi _{0}r\right) ^{2}}f_{0}(r) &=&0,\text{ \ \
\ } \\
f_{1}^{^{\prime \prime }}(r)+\frac{\omega ^{2}}{\left( \alpha -\psi
_{0}r\right) ^{2}}f_{1}(r) &=&0\text{\ \ \ \ \ \ \ } \\
f_{2}^{^{\prime \prime }}(r)+\frac{\omega \left( \omega +i\varphi
_{0}\right) }{\left( \alpha -\psi _{0}r\right) ^{2}}f_{2}(r) &=&0,
\end{eqnarray}%
Thus, the solutions in the asymptotic case are%
\begin{eqnarray}
f_{0}(r) &=&C_{1}\text{\ }\left( \alpha -\psi _{0}r\right) ^{\frac{\varphi
_{0}+i\omega }{\varphi _{0}}}+C_{2}\text{\ }\left( \alpha -\psi _{0}r\right)
^{\frac{-i\omega }{\varphi _{0}}} \\
f_{1}(r) &=&C_{3}\text{\ }\left( \alpha -\psi _{0}r\right) ^{\gamma
_{1}}+C_{4}\text{\ }\left( \alpha -\psi _{0}r\right) ^{\gamma _{1}}, \\
f_{2}(r) &=&C_{5}\text{\ }\left( \alpha -\psi _{0}r\right) ^{\frac{\varphi
_{0}-i\omega }{\varphi _{0}}}+C_{6}\text{\ }\left( \alpha -\psi _{0}r\right)
^{\frac{i\omega }{\varphi _{0}}}
\end{eqnarray}%
in which $C_{i}$ are integration constants, $\gamma _{1}=\frac{\psi _{0}+%
\sqrt{\psi _{0}^{2}-4\omega ^{2}}}{2\psi _{0}}$ and $\gamma _{2}=\frac{\psi
_{0}-\sqrt{\psi _{0}^{2}-4\omega ^{2}}}{2\psi _{0}}.$ The square
integrability condition at infinity is checked by calculating the squared
norm of each solution $f_{i}$ 
\begin{equation}
\left\Vert f_{i}\right\Vert ^{2}=\int_{r}^{\infty }\frac{\left\vert
f_{i}\left( r\right) \right\vert ^{2}r^{2}}{(\alpha -\psi _{0}r)}dr.\text{ \
\ \ \ \ \ \ }i=0,1,2
\end{equation}%
Calculations has revealed that the obtained solutions do not belong to the
Hilbert space because $\left\Vert f_{i}\right\Vert ^{2}\rightarrow \infty .$

\subsubsection{The case r$\rightarrow 0$}

The metric near $r\rightarrow 0$ is given in Eq. (47). Hence, the radial
parts of the Maxwell equations (81), (82) and (83) for this case are given by%
\begin{eqnarray}
f_{0}^{^{\prime \prime }}(r)+\frac{2}{r}f_{0}^{^{\prime }}(r)+\frac{a_{0}}{r}%
f_{0}(r) &=&0\text{, \ } \\
f_{1}^{^{\prime \prime }}(r)-\frac{1}{r}f_{1}^{^{\prime }}(r)+\frac{b_{0}}{r}%
f_{0}(r) &=&0, \\
f_{2}^{^{\prime \prime }}(r)-\frac{2}{r}f_{2}^{^{\prime }}(r)+\frac{a_{0}}{r}%
f_{0}(r) &=&0\text{ }
\end{eqnarray}%
in which $a_{0}=\frac{\epsilon ^{2}}{2GM},$ $b_{0}=\frac{\eta ^{2}}{2GM}$
and solutions are obtained as,%
\begin{eqnarray}
f_{0}(r) &=&\frac{C_{1}}{\sqrt{r}}J_{1}(2\sqrt{a_{0}r})+\frac{C_{2}}{\sqrt{r}%
}N_{1}(2\sqrt{a_{0}r}), \\
f_{1}(r) &=&C_{3}rJ_{2}(2\sqrt{b_{0}r})+C_{4}rN_{2}(2\sqrt{b_{0}r}), \\
f_{2}(r) &=&C_{5}r^{3/2}J_{3}(2\sqrt{a_{0}r})+C_{6}r^{3/2}N_{3}(2\sqrt{a_{0}r%
}),
\end{eqnarray}%
where $C_{i}$ are constants, $J_{i}$ and $N_{i}$ are Bessel and Neumann
functions. The above solutions is checked for square integrability.
Calculations have revealed that 
\begin{equation}
\left\Vert f_{i}\right\Vert ^{2}=\int_{0}^{\text{constant}}\frac{\left\vert
f_{i}\left( r\right) \right\vert ^{2}r^{2}}{\left( \alpha -\frac{2GM}{r}%
\right) }dr<\infty ,
\end{equation}%
which indicates that the obtained solutions are square integrable. As a
result, the spatial part of the Maxwell operator is not essentially
self-adjoint and therefore, the occurrence of the timelike naked singularity
in $f(R)$ gravity is quantum mechanically singular, if it is probed with
photon waves.

\subsection{Dirac Fields}

The Newman-Penrose formalism will also be used here to find the massless
Dirac fields (fermions) propagating in the space of $f(R)$ global monopole
spacetime. The Chandrasekhar-Dirac (CD) equations in the Newman-Penrose
formalism are given by

\begin{eqnarray}
\left( D+\epsilon -\rho \right) F_{1}+\left( \bar{\delta}+\pi -\alpha
\right) F_{2} &=&0, \\
\left( \Delta +\mu -\gamma \right) F_{2}+\left( \delta +\beta -\tau \right)
F_{1} &=&0, \\
\left( D+\bar{\epsilon}-\bar{\rho}\right) G_{2}-\left( \delta +\bar{\pi}-%
\bar{\alpha}\right) G_{1} &=&0, \\
\left( \Delta +\bar{\mu}-\bar{\gamma}\right) G_{1}-\left( \bar{\delta}+\bar{%
\beta}-\bar{\tau}\right) G_{2} &=&0,
\end{eqnarray}%
where $F_{1},F_{2},G_{1}$ and $G_{2}$ are the components of the wave
function, $\epsilon ,\rho ,\pi ,\alpha ,\mu ,\gamma ,\beta $ and $\tau $ are
the spin coefficients. The non-zero spin coefficients are given in Eq.s
(27,28). The directional derivatives in the CD equations are the same as in
the Maxwell equations. Substituting non-zero spin coefficients and the
definitions of the operators given in Eq.s (59-62) into the CD equations
leads to

\begin{gather}
\left( \mathbf{D}_{0}+\frac{1}{r}\right) F_{1}+\frac{1}{r\sqrt{2}}\mathbf{L}%
_{1}F_{2}=0, \\
-\frac{B}{2}\left( \mathbf{D}_{0}^{\dagger }+\frac{B^{^{\prime }}}{2B}+\frac{%
1}{r}\right) F_{2}+\frac{1}{r\sqrt{2}}\mathbf{L}_{1}^{\dagger }F_{1}=0, \\
\left( \mathbf{D}_{0}+\frac{1}{r}\right) G_{2}-\frac{1}{r\sqrt{2}}\mathbf{L}%
_{1}^{\dagger }G_{1}=0, \\
\frac{B}{2}\left( \mathbf{D}_{0}^{\dagger }+\frac{B^{^{\prime }}}{2B}+\frac{1%
}{r}\right) G_{1}+\frac{1}{r\sqrt{2}}\mathbf{L}_{1}G_{2}=0.
\end{gather}%
For the solution of the CD equations, we assume a separable solution in the
form of%
\begin{eqnarray}
F_{1} &=&f_{1}(r)Y_{1}(\theta )e^{i\left( kt+m\varphi \right) }, \\
F_{2} &=&f_{2}(r)Y_{2}(\theta )e^{i\left( kt+m\varphi \right) }, \\
G_{1} &=&g_{1}(r)Y_{3}(\theta )e^{i\left( kt+m\varphi \right) }, \\
G_{2} &=&g_{2}(r)Y_{4}(\theta )e^{i\left( kt+m\varphi \right) },
\end{eqnarray}%
where $m$ is the azimuthal quantum number and $k$ is the frequency of the
Dirac fields, which is assumed to be positive and real. Since $\left\{
f_{1},f_{2},g_{1},g_{2}\right\} $ and $\left\{
Y_{1},Y_{2},Y_{3},Y_{4}\right\} $ are functions of $r$ and $\theta ,$
respectively, by substituting Eq.s (106-109) into Eq.s (102-105) and
applying the assumptions given by%
\begin{eqnarray}
\text{\ }f_{1}(r) &=&g_{2}(r)\text{ \ \ \ \ and \ \ \ }f_{2}(r)=g_{1}(r)%
\text{\ \ }, \\
Y_{1}(\theta ) &=&Y_{3}(\theta )\text{ \ \ \ \ and \ \ \ }Y_{2}(\theta
)=Y_{4}(\theta ),
\end{eqnarray}%
the Dirac equations transform into Eq.s (112,113) below. In order to solve
the radial equations, the separation constant $\lambda $ should be defined.
This is achieved by using the angular equations. In fact, it is already
known from the literature that the separation constant can be expressed in
terms of the spin-weighted spheroidal harmonics. The radial parts of the
Dirac equations become

\begin{gather}
\left( \mathbf{D}_{0}+\frac{1}{r}\right) f_{1}\left( r\right) =\frac{\lambda 
}{r\sqrt{2}}f_{2}\left( r\right) , \\
\frac{B}{2}\left( \mathbf{D}_{0}^{\dagger }+\frac{B^{^{\prime }}}{2B}+\frac{1%
}{r}\right) f_{2}\left( r\right) =\frac{\lambda }{r\sqrt{2}}f_{1}\left(
r\right) .
\end{gather}%
We further assume that

\begin{eqnarray}
f_{1}\left( r\right) &=&\frac{\Psi _{1}\left( r\right) }{r}, \\
f_{2}\left( r\right) &=&\frac{\Psi _{2}\left( r\right) }{r},
\end{eqnarray}%
then Eq.s (112,113) transforms into,

\begin{gather}
\mathbf{D}_{0}\Psi _{1}=\frac{\lambda }{r\sqrt{2}}\Psi _{2}, \\
\frac{B}{2}\left( \mathbf{D}_{0}^{\dagger }+\frac{B^{^{\prime }}}{2B}\right)
\Psi _{2}=\frac{\lambda }{r\sqrt{2}}\Psi _{1}.
\end{gather}%
Note that $\sqrt{\frac{B}{2}}\mathbf{D}_{0}^{\dagger }\sqrt{\frac{B}{2}}=%
\mathbf{D}_{0}^{\dagger }+\frac{B^{^{\prime }}}{2B}+\frac{1}{r}$, and using
this together with the new functions%
\begin{eqnarray}
R_{1}\left( r\right)  &=&\Psi _{1}\left( r\right) , \\
R_{2}\left( r\right)  &=&\sqrt{\frac{B}{2}}\Psi _{2}\left( r\right) ,
\end{eqnarray}%
and defining the tortoise coordinate $r_{\ast }$ as%
\begin{equation}
\frac{d}{dr_{\ast }}=B\frac{d}{dr},
\end{equation}%
Eq.s (116,117) become%
\begin{eqnarray}
\left( \frac{d}{dr_{\ast }}+ik\right) R_{1} &=&\frac{\sqrt{B}\lambda }{r}%
R_{2}, \\
\left( \frac{d}{dr_{\ast }}-ik\right) R_{2} &=&\frac{\sqrt{B}\lambda }{r}%
R_{1},
\end{eqnarray}%
In order to write Eq.s (121,122) in a more compact form, we combine the
solutions in the following way:%
\begin{eqnarray}
Z_{+} &=&R_{1}+R_{2}, \\
Z_{-} &=&R_{2}-R_{1}.
\end{eqnarray}%
After doing some calculations we end up with a pair of one-dimensional Schr%
\"{o}dinger-like wave equations with effective potentials,

\begin{gather}
\left( \frac{d^{2}}{dr_{\ast }^{2}}+k^{2}\right) Z_{\pm }=V_{\pm }Z_{\pm },
\\
V_{\pm }=\left[ \frac{B\lambda ^{2}}{r^{2}}\pm \lambda \frac{d}{dr_{\ast }}%
\left( \frac{\sqrt{B}}{r}\right) \right] .
\end{gather}%
In analogy with Eq. (33), the radial operator $A$ for the Dirac equations
can be written as,

\begin{equation}
A=-\frac{d^{2}}{dr_{\ast }^{2}}+V_{\pm },
\end{equation}%
If we write the above operator in terms of the usual coordinates $r,$ by
using Eq. (120), we have

\begin{multline}
A=-\frac{d^{2}}{dr^{2}}-\frac{B^{^{\prime }}}{B}\frac{d}{dr}+ \\
\frac{\lambda }{B}\left[ \frac{\lambda }{r^{2}}\pm \frac{d}{dr}\left( \frac{%
\sqrt{B}}{r}\right) \right] ,
\end{multline}%
Our aim now is to show whether this radial part of the Dirac operator is
essentially self-adjoint or not. This will be achieved by considering Eq.
(38) and counting the number of solutions that do not belong to Hilbert
space. Hence, Eq. (38) becomes

\begin{multline}
\left( \frac{d^{2}}{dr^{2}}+\frac{B^{^{\prime }}}{B}\frac{d}{dr}-\right. \\
\left. \frac{\lambda }{B}\left[ \frac{\lambda }{r^{2}}\pm \frac{d}{dr}\left( 
\frac{\sqrt{B}}{r}\right) \right] \mp i\right) \psi (r)=0.
\end{multline}

\subsubsection{For the case $r\rightarrow \infty $}

For the asymptotic case, $r\rightarrow \infty $ , the above equation
transforms to 
\begin{equation}
\frac{d^{2}\psi \left( r\right) }{dr^{2}}\pm i\psi \left( r\right) =0,
\end{equation}%
whose solution is%
\begin{equation}
\psi _{\pm }\left( r\right) =C_{1}\sin \left[ \frac{\left( 1\pm i\right) }{%
\sqrt{2}}r\right] +C_{2}\cos \left[ \frac{\left( 1\pm i\right) }{\sqrt{2}}r%
\right]
\end{equation}%
The square integrability condition at infinity is checked by calculating the
squared norm of each sign of solution $\psi _{\pm }\left( r\right) $ 
\begin{equation}
\parallel \psi _{\pm }\left( r\right) \parallel ^{2}=\int_{r}^{\infty }\frac{%
\left\vert \psi _{\pm }\left( r\right) \right\vert ^{2}r^{2}}{(\alpha -\psi
_{0}r)}dr.\text{ \ \ \ \ \ \ \ }
\end{equation}%
The outcome of the calculations showed that the obtained solutions are not
belong to the Hilbert space because $\left\Vert \psi _{\pm }\left( r\right)
\right\Vert ^{2}\rightarrow \infty .$

\subsubsection{For the case r$\rightarrow 0$}

Near $r\rightarrow 0$ , the approximate metric is given in Eq. (47) and
hence, Eq. (129) for $r\rightarrow 0$ becomes

\begin{equation}
\frac{d^{2}\psi \left( r\right) }{dr^{2}}+\frac{i\xi }{r^{3/2}}\psi \left(
r\right) =0,
\end{equation}%
in which $\xi =\frac{\pm \lambda -2}{2\sqrt{2GM}},$ whose solution is given
by%
\begin{multline}
\psi \left( r\right) =C_{1}\left\{ -(1-i)\sqrt{2}r^{1/4}J_{1}(X)+4\sqrt{r\xi 
}J_{0}(X)\right\} + \\
C_{2}\left\{ -(1-i)\sqrt{2}r^{1/4}N_{1}(X)+4\sqrt{r\xi }N_{0}(X)\right\}
\end{multline}%
where $J_{i}\left( X\right) $ and $N_{i}\left( X\right) $ are Bessel
functions of the first and second kind, and $X=2(1+i)\sqrt{2\xi }r^{1/4}.$
Checking for the square integrability near $r\rightarrow 0$ has revealed
that both solutions are square integrable.

Hence, the radial operator of the Dirac field fails to satisfy a unique
self-adjoint extension condition for the entire space. As a result, the
occurrence of the timelike naked singularity in the context of $f(R)$ global
monopole remains singular from the quantum mechanical point of view, if it
is probed with fermions.

\section{Probing the Singularity Around BV Spacetime with Maxwell and Dirac
Fields}

In this section, we will extend the study of Pitelli and Letelier \cite{17}
for the BV spacetime in which the bosonic waves obeying the Klein-Gordon
equation is used to probe the singularity to the spinor fields obeying the
Maxwell and Dirac equations. Our motivation here is to check whether the
spinorial waves can smooth out the singularity or not. The metric describing
global monopole was obtained by BV and given by%
\begin{equation}
ds^{2}=dt^{2}-dr^{2}-a^{2}r^{2}\left( d\theta ^{2}+\sin ^{2}\theta d\varphi
^{2}\right) .
\end{equation}%
The appropriate tetrads and the non zero spin coefficients are given by%
\begin{eqnarray}
l^{a} &=&\left( 1,1,0,0\right) , \\
n^{a} &=&\left( \frac{1}{2},-\frac{1}{2},0,0\right) , \\
m^{a} &=&\frac{1}{\sqrt{2}}\left( 0,0,\frac{1}{ar},\frac{i}{ra\sin \theta }%
\right) . \\
\bar{m}^{a} &=&\frac{1}{\sqrt{2}}\left( 0,0,\frac{1}{ar},\frac{-i}{ra\sin
\theta }\right) 
\end{eqnarray}%
\begin{equation}
\mu =-\frac{1}{2r},\text{ }\rho =-\frac{1}{r},\text{ }\beta =-\alpha =\frac{1%
}{2\sqrt{2}}\frac{\cot \theta }{ra}.
\end{equation}%
The non-vanishing tetrad fields are 
\begin{equation}
\Psi _{2}=-2\Lambda =-\frac{2}{3}\phi _{11}=\frac{\left( 1-\frac{1}{a^{2}}%
\right) }{6r^{2}}
\end{equation}%
which vanish for $a=\pm 1.$

\subsection{Maxwell Fields}

Following the same steps of the previous section, the radial part of the
Maxwell's equations (51-54) governing the photon waves are obtained as%
\begin{equation}
f_{0}^{^{\prime \prime }}(r)+\left[ \omega ^{2}-\frac{2i\omega }{r}-\frac{%
\epsilon ^{2}}{r^{2}a^{2}}\right] f_{0}(r)=0,
\end{equation}%
\begin{equation}
f_{1}^{^{\prime \prime }}(r)+\left[ \omega ^{2}-\frac{\eta ^{2}}{r^{2}a^{2}}%
\right] f_{1}(r)=0,
\end{equation}%
\begin{equation}
f_{2}^{^{\prime \prime }}(r)+\left[ \omega ^{2}+\frac{2i\omega }{r}-\frac{%
\epsilon ^{2}}{r^{2}a^{2}}\right] f_{2}(r)=0.
\end{equation}

\subsubsection{For the case r$\rightarrow \infty $}

For the asymptotic case the Maxwell's equations reduces to 
\begin{equation}
f_{i}^{^{\prime \prime }}(r)+\omega ^{2}f_{i}(r)=0,\text{ \ \ }i=0,1,2
\end{equation}%
whose solution is 
\begin{equation}
f_{i}\left( r\right) =C_{1}\sin \left( \omega r\right) +C_{2}\cos \left(
\omega r\right) ,\text{ \ \ }i=0,1,2\text{\ \ }
\end{equation}%
in which $C_{1}$ and $C_{2}$ are arbitrary constants. The square
integrability condition at infinity is calculated by 
\begin{equation}
\left\Vert f_{i}\left( r\right) \right\Vert ^{2}=\int_{r}^{\infty
}\left\vert f_{i}\left( r\right) \right\vert ^{2}r^{2}dr,
\end{equation}%
and it is found that the squared norm $\left\Vert f_{i}\left( r\right)
\right\Vert ^{2}\rightarrow \infty .$ This result indicates that all the
asymptotic solutions of the Maxwell's equation do not belong to the Hilbert
space.

\subsubsection{For the case r$\rightarrow 0$}

The Maxwell's equations near $r=0$ behaves as%
\begin{equation}
f_{i}^{^{\prime \prime }}(r)-\frac{\epsilon ^{2}}{r^{2}\alpha ^{2}}%
f_{i}(r)=0,\text{ \ \ }i=0,2
\end{equation}%
\begin{equation}
f_{1}^{^{\prime \prime }}(r)-\frac{\eta ^{2}}{r^{2}\alpha ^{2}}f_{1}(r)=0.
\end{equation}%
The solutions to these equations are obtained as%
\begin{equation}
f_{i}(r)=C_{3i}r^{\gamma _{1}}+C_{4i}r^{\gamma _{2}},\text{\ \ }i=0,2\text{\ 
}
\end{equation}%
and%
\begin{equation}
f_{1}(r)=C_{5}r^{\gamma _{3}}+C_{6}r^{\gamma _{4}},
\end{equation}%
where $C_{3i},C_{4i},C_{5}$ and $C_{6}$ are arbitrary constants. The
exponents are given by%
\begin{eqnarray}
\gamma _{1} &=&\frac{1}{2}\left( 1+\sqrt{1+\frac{4\epsilon ^{2}}{a^{2}}}%
\right) ,\text{ \ \ } \\
\gamma _{2} &=&\frac{1}{2}\left( 1-\sqrt{1+\frac{4\epsilon ^{2}}{a^{2}}}%
\right) , \\
\gamma _{3} &=&\frac{1}{2}\left( 1+\sqrt{1+\frac{4\eta ^{2}}{a^{2}}}\right) ,%
\text{ \ } \\
\text{\ }\gamma _{4} &=&\frac{1}{2}\left( 1-\sqrt{1+\frac{4\eta ^{2}}{a^{2}}}%
\right) .
\end{eqnarray}%
The square integrability near $r=0$ is checked by calculating the squared
norms of the obtained solutions by%
\begin{equation}
\parallel f_{i}\parallel ^{2}=\int_{0}^{\text{constant}}\left\vert
f_{i}\left( r\right) \right\vert ^{2}r^{2}dr.
\end{equation}%
Our analysis has revealed that, if $C_{3i}=C_{5}=0$ together with $\frac{%
\epsilon ^{2}}{\alpha ^{2}}>\frac{15}{4}$ and $\frac{\eta ^{2}}{\alpha ^{2}}>%
\frac{15}{4}$, the squared norms diverges. This result implies that the
solutions for these specific modes do not belong to the Hilbert space.

Consequently, in contrast to the bosonic wave probe reported in \cite{17},
the classical singularity at $r=0,$ for global monopole spacetime due to the
BV, remains quantum mechanically nonsingular with respect to the photonic
wave probe that has spin 1.

\subsection{Dirac Fields}

The Chandrasekhar-Dirac equations given in Eq.s (98-101) is solved by using
the Newman-Penrose formalism for the ordinary global monopole metric (135).
The same steps are followed as in section IV and hence, we end up with a
pair of one - dimensional Schr\"{o}dinger-like wave equations with effective
potentials, 
\begin{equation}
\left( \frac{d^{2}}{dr^{2}}+k^{2}\right) Z_{\pm }=V_{\pm }Z_{\pm },
\end{equation}

\begin{equation}
V_{\pm }=\frac{\lambda ^{^{\prime }2}}{r^{2}}\mp \frac{\lambda ^{^{\prime }}%
}{r^{2}}.
\end{equation}%
in which $\lambda ^{^{\prime }}=\frac{\lambda }{\alpha }.$ Comparing with
the equation (33), the radial operator $A$ for the Dirac equations can be
written as%
\begin{equation}
A=-\frac{d^{2}}{dr^{2}}+V_{\pm }.
\end{equation}%
As a requirement of the HM criterion, the radial Dirac operator $A$ should
be examined whether it is essentially self-adjoint or not. We obtain this by
considering Eq. (38) and counting the number of solutions for each sign that
do not belong to Hilbert space. Hence, we have%
\begin{equation}
\left( \frac{d^{2}}{dr^{2}}-\left[ \frac{\lambda ^{^{\prime }2}}{r^{2}}\mp 
\frac{\lambda ^{^{\prime }}}{r^{2}}\right] \mp i\right) \psi (r)=0.
\end{equation}

\subsubsection{For the case r$\rightarrow \infty $}

The behavior of the Eq. (160), as $r\rightarrow \infty $ is%
\begin{equation}
\left( \frac{d^{2}}{dr^{2}}\mp i\right) \psi (r)=0,
\end{equation}%
whose solutions for each sign is 
\begin{multline}
\psi _{\pm }(r)=C_{1\pm }\sin \left( \frac{1}{\sqrt{2}}\left( 1\pm i\right)
r\right) + \\
C_{2\pm }\cos \left( \frac{1}{\sqrt{2}}\left( 1\pm i\right) r\right) ,
\end{multline}%
in which $C_{1\pm }$ and $C_{2\pm }$ are arbitrary integration constants for
each sign of solution. Our calculations has shown that, the squared norms
for each sign of solutions diverges, that is%
\begin{equation}
\left\Vert \psi _{\pm }\left( r\right) \right\Vert ^{2}=\int_{r}^{\infty
}\left\vert \psi _{\pm }\left( r\right) \right\vert ^{2}r^{2}dr\rightarrow
\infty ,
\end{equation}%
indicating that the solutions at infinity do not belong to the Hilbert space.

\subsubsection{For the case r$\rightarrow 0$}

The behavior of the Eq. (160), near $r=0$ is,%
\begin{equation}
\left( \frac{d^{2}}{dr^{2}}-\frac{\lambda ^{^{\prime }}}{r^{2}}\left[
\lambda ^{^{\prime }}\mp 1\right] \right) \psi (r)=0.
\end{equation}%
The solution is 
\begin{equation}
\psi (r)=C_{3}r^{\tau _{1}}+C_{4}r^{\tau _{2}},
\end{equation}%
in which $C_{3}$ and $C_{4}$ are arbitrary constants. The exponents are
given by%
\begin{eqnarray}
\tau _{1} &=&\frac{1}{2}\left( 1+\sqrt{1+4\lambda ^{^{\prime }}\left(
\lambda ^{^{\prime }}\pm 1\right) }\right) ,\text{ \ \ \ \ } \\
\text{\ }\tau _{1} &=&\frac{1}{2}\left( 1-\sqrt{1+4\lambda ^{^{\prime
}}\left( \lambda ^{^{\prime }}\pm 1\right) }\right) .
\end{eqnarray}%
The obtained solution fails to be square integrable, if $C_{3}=0$ and $%
\lambda ^{^{\prime }}\left( \lambda ^{^{\prime }}\pm 1\right) >\frac{15}{4}.$
Hence, solutions for these modes do not belong to the Hilbert space. As a
result, the classical singularity at $r=0$, remains quantum mechanically
nonsingular, if it is probed with fermions whose spin structure is $1/2$.

\section{Analysis with Gubser's Singularity Conjecture}

In this section, Gubser's \cite{GUB} singularity conjecture will be used to
analyse the timelike naked singularity in the $f(R)$ global monopole
spacetime. It should be noted that this conjecture is based on investigating
the behavior of the scalar potential $V(\overrightarrow{\varphi })$ on
shell. Hence, the Gubser's singularity conjecture, investigates the
singularity from a geometric point of view. Apparently different but
structurally equivalent singularity criteria are proposed by Kim in \cite{KM}%
, in which $D+1$ dimensional geometry with $D$ Poincare invariant spacetime
is considered in the following form, 
\begin{equation}
ds^{2}=a\left( y\right) ^{2}\eta _{\mu \nu }dx^{\mu }dx^{\nu }+dy^{2}.
\end{equation}%
It is argued that, if the integral of the on-shell Lagrangian density over
the finite range of $y$, whose least upper bound is $y=y_{c}$ , is finite,
the singularity at $y=y_{c}$ is physically admissible.

The $4-$dimensional global monopole spacetime is governed by the triplet
scalar field coupled with gravity. Therefore, we believe that this
conjecture is applicable in this theory too. We consider the simplest case
in which the action is given by%
\begin{equation}
I=\int d^{4}x\sqrt{-g}\left( R+\frac{1}{2}\partial _{\mu }\phi ^{a}\partial
^{\mu }\phi ^{a}+V\left( \phi ^{a}\right) \right)
\end{equation}%
where $V\left( \phi ^{a}\right) =-\frac{1}{4}\lambda \left( \phi ^{a}\phi
^{a}-\eta ^{2}\right) ^{2}$ and $a=1,2,3$ \cite{27}. It is clear that $%
V\left( \phi ^{a}\right) $ has a local maximum at $\phi ^{a}\phi ^{a}=\eta
^{2}.$ Fig. 4 shows a contour plot of $V\left( \phi ^{a}\right) $ with
respect to $\phi ^{1}$ and $\phi ^{2}$ while $\left( \phi ^{3}\right)
^{2}-\eta ^{2}=-\frac{1}{4}$ and $\lambda =1.$ The corresponding
superpotential \cite{29} $W\left( \phi ^{a}\right) =-\frac{\sqrt{3}}{2}%
\left( \left( \phi ^{1}\right) ^{2}+\left( \phi ^{2}\right) ^{2}+\frac{1}{4}%
\right) $ is also plotted in Fig. 5. Based on these figures one concludes
that according to the Gubser's conjecture the singularity of this spacetime
is admissible i.e. a 'good' one. 
\begin{figure}[tbp]
\includegraphics[width=60mm,scale=0.7]{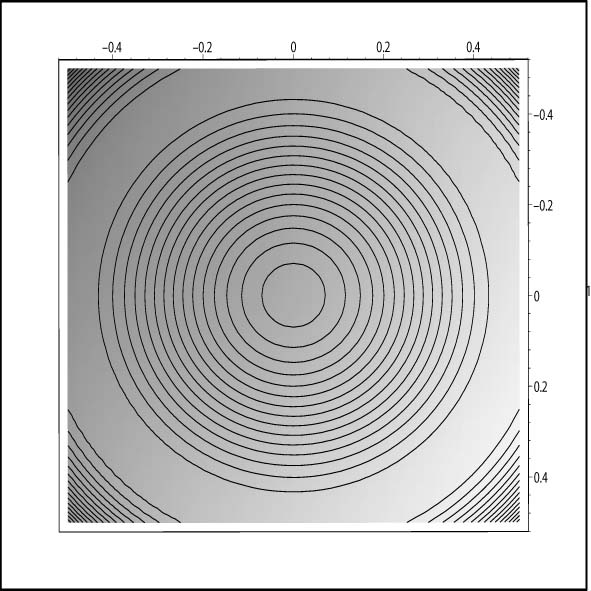}
\caption{Contour plot of $V\left( \protect\phi ^{a}\right) $ with respect to 
$\protect\phi ^{1}$ and $\protect\phi ^{2}$ when $\left( \protect\phi %
^{3}\right) ^{2}-\protect\eta ^{2}=-\frac{1}{4}$ and $\protect\lambda =1.$ A
local / absolute maximum is observed at $\left( \protect\phi ^{1}\right)
^{2}+\left( \protect\phi ^{2}\right) ^{2}=\frac{1}{4}$ and therefore the
potential is bounded from above. From the Gubser's conjecture the large
scalar curvature in this spacetime is allowed.}
\end{figure}
\begin{figure}[tbp]
\includegraphics[width=60mm,scale=0.7]{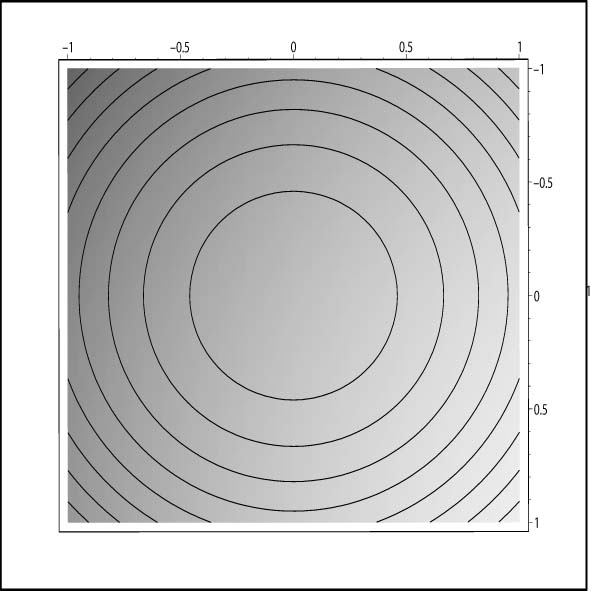}
\caption{Contour plot of $W\left( \protect\phi ^{a}\right) $ with respect to 
$\protect\phi ^{1}$ and $\protect\phi ^{2}$ when $\left( \protect\phi %
^{3}\right) ^{2}-\protect\eta ^{2}=-\frac{1}{4}$ and $\protect\lambda =1.$ A
local / absolute maximum is observed at $\left( \protect\phi ^{1}\right)
=\left( \protect\phi ^{2}\right) =0$ and therefore the superpotential is
bounded from above. }
\end{figure}

\section{Conclusion}

In this paper, the formation of the timelike naked singularity in $f(R)$
global monopole spacetime is investigated within the framework of quantum
mechanics. The timelike naked singularity developed at $r=0,$ is probed with
the quantum fields obeying the Klein$-$Gordon, Maxwell and Dirac equations.
Our investigation is based on the criterion proposed by HM that incorporates
the essential self-adjointness of the spatial part of the wave operator $A$
in the natural Hilbert space of quantum mechanics which is a linear function
space with square integrability.

In this paper, the spinorial field generalization of the quantum singularity
analysis of the BV spacetime reported in \cite{17} is also studied. In order
to show the influence of the modified theory on the singularity structure,
we compare the results of the standard general relativity and $f(R)$ theory.

We showed with explicit calculations that the naked singularity at $r=0,$
for the $f(R)$ global monopole spacetime, remains quantum mechanically
singular when it is probed with quantum fields having different spin
structures obeying Klein-Gordon, Maxwell and Dirac equations. It should be
noted that in the analysis of $f(R)$ global monopole; although the mass term
vanishes for large values of $r$ as in the case of BV spacetime, \ unlike
the case in BV, the mass term becomes effective for $r\rightarrow 0.$
Because of this nature, the singularity at $r=0$ becomes very strong in such
a way that irrespective of the spin structure of the fields used to probe
the singularity, the $f(R)$ global monopole spacetime remains quantum
mechanically singular.

An interesting result is obtained for the spinorial wave probe
generalization of the BV spacetime considered in \cite{17}. We proved that
for specific modes of solutions of the Maxwell and Dirac equations, the
singularity at $r=0$ is smoothed out. The main reason of this result, seems
to be the absence of the mass term. In addition, briefly we considered the
geometrical approach of Gubser \cite{29} to singularities in the present
problem of cosmic string singularity.

It will be interesting for future research to extend the quantum singularity
analysis in other $f(R)$ gravity models. Furthermore, it will be a great
achievement if the criterion proposed by HM is extended to stationary
metrics. Although the preliminary work in this direction is considered in 
\cite{KM}, the formulation has not been fully completed.

\bigskip


\begin{thebibliography}{99}
\bibitem{1} G. F. R. Ellis and B. G. Schmidt, \textit{Gen. Rel. Grav}. 
\textbf{8}, 915 (1977).

\bibitem{BS} P. Bell and P. Szekeres, \textit{Gen. Rel. Grav}. \textbf{5},
275 (1974).

\bibitem{2} G. T. Horowitz, \textit{New J. Phys.} \textbf{7}, 201 (2005).

\bibitem{3} M. Natsume, arXiv:gr-qc/0108059.

\bibitem{4} A. Ashtekar, \textit{J. Phys. Conf. Ser.} \textbf{189}, 012003
(2009).

\bibitem{5} J. Polchinski, \textit{String theory, Cambridge University
Press, Cambridge} 1998.

\bibitem{amit} A.Giveon, B. Kol, A. Sever and A. Ori, \textit{JHEP}, 08, 014
(2004).

\bibitem{6} R. M. Wald, \textit{J. Math. Phys. (N.Y.)} \textbf{21}, 2082
(1980).

\bibitem{7} G. T. Horowitz and D. Marolf, \textit{Phys. Rev. D} \textbf{52},
5670 (1995).

\bibitem{8} A. Ishibashi and A. Hosoya, \textit{Phys. Rev. D} \textbf{60},
104028 (1999).

\bibitem{9} D. A. Konkowski and T. M. Helliwell, \textit{Gen. Rel. and Grav.}
\textbf{33}, 1131, (2001).

\bibitem{10} T. M. Helliwell, D. A. Konkowski and V. Arndt, \textit{Gen.
Rel. and Grav.} \textbf{35}, 79, (2003).

\bibitem{11} D. A. Konkowski, T. M. Helliwell and C. Wieland, \textit{Class.
Quantum Grav.} \textbf{21,} 265 (2004).

\bibitem{12} D. A. Konkowski, C. Reese, T. M. Helliwell and C. Wieland, "
Classical and Quantum Singularities of Levi-Civita Spacetimes with and
without a Cosmological Constant", in Procedings of the Workshop on the
Dynamics and Thermodynamics of Black holes and Naked Singularities, ed.
L.Fatibene, M. Francaviglia, R. Giambo and G. Megli, 2004.

\bibitem{13} D. A. Konkowski and T. M. Helliwell, \textit{Int. J. Mod. Phys.
A}, Vol.\textbf{26}, No.22, 3878-3888 (2011).

\bibitem{14} T. M. Helliwell and D. A. Konkowski, \textit{Phys, Rev. D} 
\textbf{87}, 104041 (2013).

\bibitem{15} J. P. M. Pitelli and P. S. Letelier, \textit{J. Math. Phys.} 
\textbf{\ 48,} 092501, (2007).

\bibitem{16} J. P. M. Pitelli and P. S. Letelier,\textit{\ Phys. Rev. D} 
\textbf{77,} 124030 (2008).

\bibitem{17} J. P. M. Pitelli and P. S. Letelier, \textit{Phys. Rev. D} 
\textbf{80,} 104035 (2009).

\bibitem{18} P. S. Letelier and J. P. M. Pitelli, \textit{Phys. Rev. D} 
\textbf{82,} 104046 (2010).

\bibitem{19} O. Unver and O. Gurtug, \textit{Phys. Rev. D }\textbf{82,}
084016 (2010).

\bibitem{20} S. H. Mazharimousavi, O. Gurtug and M. Halilsoy, \textit{Int.
J. Mod. Phys. D} \textbf{18, } 2061-2082, (2009).

\bibitem{21} S. H. Mazharimousavi, M. Halilsoy, I. Sakalli and O. Gurtug, 
\textit{Class. Quant. Grav}. \textbf{27}, 105005, (2010).

\bibitem{22} S. H. Mazharimousavi, O. Gurtug, M. Halilsoy and O. Unver, 
\textit{Phys. Rev. D} \textbf{84,} 124021 (2011).

\bibitem{23} O. Gurtug and T. Tahamtan, \textit{Eur. Phys. J. C} \textbf{72}%
, 2091 (2012).

\bibitem{24} T. R. P. Carames, E. R. B. de Mello and M. E. X. Guimaraes, 
\textit{Int. J. Mod. Phys: Conference Series}, \textbf{V 03}, 446-454 (2011).

\bibitem{25} M. Barriola and A. Vilenkin, \textit{Phys. Rev. Lett}. \textbf{%
63}, 341 (1989).

\bibitem{26} J. Man and H. Cheng, \textit{Phys. Rev. D} \textbf{87}, 044002
(2013).

\bibitem{GUB} S. S. Gubser, \textit{Adv. Theor. Math. Phys}. \textbf{4},
679-745, (2000).

\bibitem{27} M. Reed and B. Simon, \textit{Functional Analysis}, (Academic
Press, New York, 1980).

\bibitem{28} M. Reed and B. Simon, \textit{Fourier Analysis and
Self-Adjointness}, (Academic Press, New York, 1975).

\bibitem{29} R. D. Richtmyer, \textit{Principles of Advanced Mathematical
Physics}, (Springer, New York, 1978).

\bibitem{30} S. Chandrasekhar, \textit{The Mathematical Theory of Black Holes%
}, (Oxford University Press, 1992).

\bibitem{31} I. Seggev, \textit{Class. Quant. Grav}. \textbf{21}, 2651,
(2004).

\bibitem{KM} H. D. Kim, \textit{Phys. Rev. D} \textbf{63}, 124001, (2001).
\end{thebibliography}
\end{document}